\documentstyle[12pt]{article}
\def\d{\partial}

\def\th{\theta}
\def\g{\gamma}

\def\r{\rho}
\def\l{\lambda}

\def\L{\Lambda}

\def\s{\sigma}

\def\e{\epsilon}

\def\t{\tilde}

\def\is{\equiv}

\newfont{\ball}{eurb10}
\newcommand\pmb[1]{\mbox{\ball #1}}

\def\half{{\textstyle{1 \over 2}}}

\def\bop#1{\setbox0=\hbox{$#1M$}\mkern1.5mu
        \vbox{\hrule height0pt depth.04\ht0
        \hbox{\vrule width.04\ht0 height.9\ht0 \kern.9\ht0
        \vrule width.04\ht0}\hrule height.04\ht0}\mkern1.5mu}

\newcommand\hv[1]{(\ref{#1})}

\begin{document}

\newcommand{\inv}[1]{{#1}^{-1}} %inverse

\renewcommand{\theequation}{\thesection.\arabic{equation}}
\newcommand{\beq}{\begin{equation}}
\newcommand{\eeq}[1]{\label{#1}\end{equation}}
\newcommand{\ber}{\begin{eqnarray}}
\newcommand{\eer}[1]{\label{#1}\end{eqnarray}}
%\begin{titlepage}
\begin{center}
        		\hfill    USITP-97-03\\
			\hfill    April, 1997\\
			  
				\hfill    hep-th/9704051\\

\vskip .3in \noindent

\vskip .1in

{\large \bf A Picture of D-branes at Strong Coupling}
\vskip .2in

{\bf Ulf Lindstr\"om}$^a$\footnotemark and 
{\bf Rikard von Unge}$^b$\footnotemark \\

\footnotetext{e-mail address: ul@physto.se}
\vskip .15in

\footnotetext{e-mail address: unge@vanosf.physto.se}
\vskip .15in
$\mbox{}^{a,b)}$ {\em  Institute of Theoretical Physics,
University of Stockholm \\
Box 6730,
S-113 85 Stockholm SWEDEN}\\
\bigskip

$\mbox{}^{a)}$ {\em Institute of Physics, University of Oslo\\
Box 1048, N-0316 Blindern, Oslo, NORWAY\\}
\vskip .15in

\vskip .1in
\end{center}
\vskip .4in
\begin{center} {\bf ABSTRACT } \end{center}
\begin{quotation}\noindent
We use a phase space description to (re)derive a first order
form  of the Born-Infeld action for
$D$-branes. This derivation also makes it possible to consider the
limit where the tension of the $D$-brane goes to zero. We find that in this
limit, which can be considered to be the strong coupling limit of the
fundamental string theory, the world-volume of the
$D$-brane generically splits into a collection of tensile strings.
\end{quotation}
\vfill
\eject
\section{Introduction}
$D$-branes \cite{pol} provide a description of a certain class of string
theory solitons. They are BPS states so they are thought to survive
in the strong coupling limit. Formally one may investigate what
happens with the Born-Infeld description of $D$-branes in this
limit. Since the tension of a $D$ $p$-brane is given by the formula
\beq
 T_{p} = \frac{2\pi}{g\left(2\pi\sqrt{\alpha^{\prime}}\right)^{p+1}},
\eeq{sil1}
where $\alpha^{\prime}$ is the inverse of the fundamental string
tension and $g$ is the dimensionless string coupling $g =
e^{\phi}$, the tension of the $D$ $p$-brane goes to zero whenever
$g\rightarrow\infty$, i.e. when the fundamental string theory goes to
strong coupling\footnote{In principle the tension of the $D$-brane
goes to zero also when the fundamental string tension goes to
zero. However, the non-trivial dependence of $\alpha^{\prime}$ of the
Born-Infeld action makes this limit much more obscure.  We will
subsequently put $\alpha^{\prime}=\frac{1}{2\pi}$.}. We thus find that
the strong coupling limit is equivalent to the limit where the
$D$-brane tension goes to zero.

In this paper we rederive a first order form of the tensile
Born-Infeld action, first presented in \cite{ul}, using a Hamiltonian
approach. In the process of this derivation it becomes clear how to
take the $T\to 0$ limit and we thus find a tensionless version of the
Born-Infeld action. Somewhat surprisingly, what we find is that when we
let the tension of the $D$-brane go to zero the world volume of the
brane generically splits into a collection of tensile strings or, in
special cases, massless particles, thus leading to a parton picture of
$D$-branes in this limit.

The techniques used when taking the $T\rightarrow 0$ limit in this
paper are those previously employed in taking this limit in
fundamental string theory \cite{tless,kl,lst1,lst2,lr,sadri}, i.e. in 
constructing tensionless strings\footnote{Recently, tensionless
strings have also been considered in the context of $D$-branes
\cite{strom}-\cite{sixten}. They arise, e.g., in the compactification
of type-IIB to 6D on a $K_3$ when the area of the 2-cycle shrinks to
zero size \cite{witt1}.  In this latter context one does not know the
dynamics as yet, and the precise relation to the high energy
tensionless strings thus remains unclear.}

The plan of the paper is as follows:
We will first give a summary of our results, then
present their derivation and finally discuss the connection to
previous work.

\section{Results}
Some time ago one of the authors showed how to write a first order
version of the Born-Infeld action for a bosonic {\em p}-brane, 
\beq
S^2_{BI}(T_{p})=T_{p}\int d^{p+1}\xi \sqrt{-\det(
       \g_{ij}+F_{ij})},
\eeq{binf}
where $\g _{ij}\equiv \d _iX^\mu \d_jX^\nu G_{\mu\nu}(X)$ is the
induced metric on the world-sheet from a background metric $G_{\mu\nu}$ and 
$F_{ij}$ is a world-volume field strength, which in this article we will take 
to be
\beq
F_{ij}\equiv \d_{[i}A_{j]}+B_{ij} .
\eeq{fdef}
Here $B_{ij}\equiv\d_iX^\mu\d_jX^\nu B_{\mu\nu}$ is the pull-back of the back
ground Kalb-Ramond field. The
first order action is \cite{ul}: 
\beq 
S^1_{BI}(T)=\half T\int
d^{p+1}\xi \sqrt{-s}\left( s^{ij}(\g _{ij}+F_{ij})-(p-1)\right),
\eeq{sbinf} 
where $s^{ij}$ is a general second rank world-volume
tensor, (no symmetry as\-sumed)\footnote{This form of the action has recently
been used by Hull and Abou Zeid to discuss the geometry of $D$-branes
\cite{Hull}.}. We take \hv{sbinf} to represent the $D$ brane action,
thus disregarding the over all dilaton factor $e^{-\phi}$ in the
Lagrangian, since it will play no role for our considerations. If one
wishes, it is
readily reinserted in all our formulae.  

In this letter we derive (\ref{sbinf}) from (\ref{binf}) via a Hamiltonian
formulation. This process also allows us to take the limit $T\to 0$ and thus
obtain a formulation of the tensionless $D$-brane action, which reads 
\beq
S^1_{BI}(0)={\textstyle{1\over 4}}\int d^{p+1}\xi
(E^i_1E^j_1-E^i_2E^j_2-E^i_{[1}E^j_{2]})(\g_{ij}+F_{ij}).
\eeq{Eac}
Integrating out the $A_i$-field we are left with
\beq
\t S^1_{BI}(0)={\textstyle{1\over 4}}\int d^{p+1}\xi
\left(E_AX^\mu E_BX^\nu (\eta^{AB}G_{\mu\nu}+\e^{AB}B_{\mu\nu})\right),
\eeq{EBac}
where $A,B=0,1$, $E_A\is E^i_A\d_i$ are ``degenerate'', (for $p>1$), zwei-beins
corresponding to a $2D$ Lorentzian ``tangent space''.
In that tangent space the Minkowski metric is
$\eta^{AB}$ and $\e^{AB}$ is the epsilon symbol. 
The form of the actions \hv{Eac} and (\ref{EBac}) clearly show the nature 
of
the strong coupling limit of $D$-branes that we are considering: {\em In this 
limit
the
$D$-brane dynamics is governed by actions that involve a degenerate metric of 
rank
$2$ ,
($\propto E^i_AE^j_B\eta^{AB}$).These actions look like tensile string
actions.} We may thus expect the dynamics of these objects to be given by a
``parton'' picture with  strings as the
partons. It is very reminicent of the
tensionless string which in a particular gauge is seen to describe a collection
of massless particles moving under a certain constraint \cite{kl}. 

The parton picture will be verified in Section $5$, but let us corroborate it a
bit here. We want to interpret $E^i_AE^j_B\eta^{AB}$ as a metric
density $g^{ij}$ 
for $i,j=0,1$. This is possible since the rank of $E^i_AE^j_B\eta^{AB}$ is $2$.
Using $M,N=0,1$ to denote the two lowest values of $i,j$  we thus have
{det}$(g^{MN})$
$\is g^{-1}
\ne 0$. To make contact with the tensile string we want to write $g^{MN} \is
T\sqrt{-\t{g}}\t{g}^{MN}$ with the tension $T$ independent of
$\xi ^i$, $i\ne 0,1$, and 
$\t{g}^{MN}$ an ordinary contravariant metric (from the
$2D$ point of view). Clearly this requires $g^{-1}=-T^2$ which means that this is as
far as we get without a gauge choice. Also, we must of course make sure that the
interpretation is consistent with all the field equations.

As indicated above, the case $p=1$ is a bit special. We will treat that case
here while we defer the general case to Section $5$.

Suprisingly, for $p=1$, the action
\hv{Eac} gives a unified description of  tensile and
tensionless {\em ordinary} strings. 

We rewrite \hv{Eac} as 
\beq
\t S^1_{BI}(0)=\half\int d^{2}\xi\left(
\pmb{g}^{ij}\g_{ij}+\sqrt{-\pmb{g}}\e^{ij}F_{ij})\right),
\eeq{01Dbr}
where $\pmb{g}^{ij}$ is a symmetric second rank tensor density of
weight\footnote{In conventions where $\sqrt{-g}$ has 
weight $\half$.}
$\half$  and
$\pmb{g}$ is its determinant.
The $A_i$ field equations read 
\beq
\e^{ij}\d_j\pmb{g}=0,
\eeq{Aeq}
with solution $\pmb{g}=-\t {T}^2$, with $\t T$ a constant of dimension
$(length)^{-1}$, thus introducing a {\em new} scale in the theory. Using this we
define
$\t{T} {\t{\pmb{g}}^{ij}}\equiv \pmb{g}^{ij}$. Since $\det{
\t{\pmb{g}}^{ij}}=-1$, we have
${\t{\pmb{g}}^{ij}}=(\sqrt{-\det g_{kl}})g^{ij}$, with ${g}^{ij}$ an ordinary
metric. Plugging all this back into the action \hv{01Dbr} yields
\beq
S^1_{F}(\t T)=-\half\t T\int d^{2}\xi\left(
\sqrt{-g}{g}^{ij}\g_{ij}+\e^{ij}B_{ij}\right),
\eeq{1Fbr}
i.e., the action for a tensile (fundamental) string with string
tension $\t T$ in the $G$, $B$ background. (Note that the coupling to
the $B$-field comes out correct).

For the special case when $\t T=0$ in the solution to \hv{Aeq}, the tensor
density  becomes degenerate and may be written as $\pmb{g}^{ij}=V^iV^j$,
with $V^i$ being contravariant vector desity fields of weight $1\over 4$. This
leads to the action for a tensionless (fundamental) string \cite{lst1}:
\beq
S^1_{F}(0)=-\half\int d^{2}\xi\left(
V^iV^j\g_{ij}\right).
\eeq{10Fbr}

We conclude this section by noting that the unified description of tensile and
tensionless strings just described resembles that of \cite{blt},
although their starting point is a different formulation of the tensionless
string coupled to world-sheet electromagnetism. As will be seen in Section 4, we
recover their formulation integrating out some of the Lagrange multipliers.

\section{Derivation}
In this section we present the derivation of \hv{sbinf} from \hv{binf} for
arbitrary $p$. We follow the procedure described, e.g., in \cite{lst1}, i.e.,
derive the momenta, the constraints and then the Hamiltonian. integrating out
the momenta 
from the phase space Lagrangian we then obtain a configuration space action with
the lagrange multipliers for the constraints among the variables. We finally
identify those multipliers with geometric objects on the world volume.

The generalized momenta that follow from \hv{binf} are
\ber
\Pi_{\mu} &=& \frac{T}{2}\sqrt{-\det(\gamma+F)}
  \left[\left(\gamma+F\right)^{-1}\right]^{(0i)}
    \partial_{i}X_{\mu},\cr
P^{a} &=& \frac{T}{2}\sqrt{-\det(\gamma + F)}
  \left[\left(\gamma + F\right)^{-1}\right]^{[a0]},\cr
P^{0} &=& 0.
\eer{mom}
Using the relations
\ber
\left[\left(\gamma + F\right)^{-1}\right]^{(ik)}F_{kl} + 
  \left[\left(\gamma + F\right)^{-1}\right]^{[ik]}
  \gamma_{kl}&=& 0,\cr
\left[\left(\gamma + F\right)^{-1}\right]^{(ik)}\gamma_{kl} 
 + \left[\left(\gamma + F\right)^{-1}\right]^{[ik]}
  F_{kl} &=& 2 \delta ^{i}_{l},
\eer{rel}
one finds the primary constraints
\ber
 \Pi_{\mu}\partial_{a} X^{\mu} + P^{b}F_{ab} &=& 0,\cr
 P^{0} &=& 0,\cr
 \Pi_{\mu}\Pi^{\mu} + P^{a}\gamma_{ab}P^{b} + T^2\det\left(\left(\gamma+F
   \right)_{ab}\right) &=& 0,
\eer{con1}
where $a,b,...=1,...,p$ are transversal indices.
The corresponding Hamiltonian is 
\ber
 {\cal H} &=&
 P^{i}\partial_{i} A_{0} + \sigma P^{0} +\rho^{a}\left(\Pi_{\mu}\partial_{a}
  X^{\mu} + P^{b}F_{ab}\right) \cr
  && +\lambda \left( \Pi_{\mu}\Pi^{\mu}
  +P^{a}\gamma_{ab}P^{b} + T^2\det\left(\left(\gamma+F\right)_{ab}\right)
   \right),
\eer{h1}
where $\s ,\r^a$, and $\l $ are Lagrange multiplers for the constraints.
Preservation of the constraints \hv{con1} leads to the secondary "Gauss' law"
type constraint 
\beq
 \partial_{i}P^{i} = 0,
\eeq{gauss}
but no tertiary constraints.
The final Hamiltonian is just the sum of the
constraints, in agreement with the diffeomorphism invariance of the original
Lagrangian.
The phase space action is
\ber
S_{PS} &=& \int d^{p+1}\xi \left[\Pi_{\mu}\partial_{0}X^{\mu} \right. 
  + \left. P^{a}F_{0a} - \sigma
P^{0}\right.\cr
  &&- \lambda\left(\Pi_{\mu}\Pi^{\mu} + P^{a}\gamma_{ab}P^{b} + 
     T^2\det\left(\left(\gamma+F\right)_{ab}\right)\right)\cr
 &&- \left.\rho^{a}\left(\Pi_{\mu}\partial_{a}X^{\mu} + P^{b}F_{ab}
   \right)\right].
\eer{sps}
Integrating out the momenta yields the configuration space action
\ber
S_{CS} &=& \int d^{p+1}\xi  \frac{1}{4\lambda}\left[\gamma_{00}
-2\rho^{a}\gamma_{0a}
  +\rho^{a}\rho^{b}\gamma_{ab} \right. \cr
  &&+ \left. \left(\hat{\gamma}^{-1}\right)^{ab}
  \left(F_{0a}-\rho^{c}F_{ca}\right)\left(F_{0b}-\rho^{d}F_{db}\right)
 \right. \cr
&&\left.
-4\lambda^{2}T^{2}\det\left(\left(\gamma+F\right)_{ab}\right)\right],
\eer{scs}
where $\hat{\gamma}_{ab}$ is the spacial part of the induced metric
$\hat{\gamma}_{ab}=\gamma_{ab}$.
To linearize these realtions we introduce additional lagrange multipliers
$G^{a},\theta^{a},\sigma^{ab},H_{ab}$ and write the action as
\ber
 S &=& \int d^{p+1}\xi\frac{1}{4\lambda}\left(\gamma_{00} -2\rho^{a}\gamma_{0a}
  +\rho^{a}\rho^{b}\gamma_{ab} + G^{a}\gamma_{ab}G^{b}
  \right.\cr
 &&+\left.
  2\theta^{a}\left(\left(F_{0a}-\rho^{b}F_{ba}\right)-G^{b}\gamma_{ba}\right)
  \right. \cr
  &&-\left. 4\lambda^{2}T^2\left(\det H_{ab} + \sigma^{ab}\left(H_{ab}-F_{ab}
  -\gamma_{ab}\right)\right)\right).
\eer{act}
If we integrate out $\theta^{a},\sigma^{ab}$ we get back to the original model
but if we integrate out $G^{a}, H_{ab}$ we get
\begin{eqnarray}\label{slag}
S &=& \int d^{p+1}\xi\frac{1}{4\lambda}\left(\gamma_{00} - \rho^{a}\gamma_{0a}
  +\theta^{a}F_{0a} - \rho^{a}\gamma_{a0} - \theta^{a}F_{a0}
 \right. \nonumber\\ 
  &&+\left. \left(\rho^{a}\rho^{b} - \theta^{a}\theta^{b} + 4\lambda^{2}T^{2}
  \sigma^{ab}\right)\gamma_{ab}
 \right. \\
  &&+\left. \left(\theta^{a}\rho^{b}-\theta^{b}\rho^{a} + 4\lambda^{2}T^{2}
  \sigma^{ab}\right)F_{ab}
  +4\lambda^2T^2(p-1)H\right),\nonumber
\end{eqnarray}
where $H$ is implicitly defined in terms of $\sigma^{ab}$ through the
relation 
\ber
 \sigma^{ab}= -\det\left( H\right) \left(H^{-1}\right)^{ba}.
\eer{H}
With the identification
\ber
 s^{00} &=& -\frac{1}{4\lambda^2T^2 H},\cr
 s^{0a} &=& \frac{\rho^{a}-\theta^{a}}{4\lambda^{2}T^{2}H},\cr
 s^{a0} &=& \frac{\rho^{a}+\theta^{a}}{4\lambda^{2}T^{2}H},\cr
 s^{ab} &=& -\frac{\left(\rho^{a}+\theta^{a}\right)
    \left(\rho^{b}-\theta^{b}\right)}{4\lambda^{2}T^{2}H} - 
    \frac{\sigma^{ab}}{H},
\eer{q}
we may finally write the action as
\ber
 S = -\frac{T}{2}\int d^{p+1}\xi \sqrt{-s}\left[ s^{ik}\left(
  \partial_{i}X^{\mu}\partial_{k}X_{\mu} + F_{ik}\right) - \left(p-1\right)
  \right].
\eer{fin}

\section{The tensionless theory}
The action in (\ref{slag}) is suitable for taking the tension to
zero. Doing that we obtain the action
\ber
S &=& \int d^{p+1}\xi\frac{1}{4\lambda} 
 \left(\gamma_{00} - 2\rho^{a}\gamma_{0a}
  +2\theta^{a}F_{0a}
 \right. \nonumber\\ && \left.
  +\left(\rho^{a}\rho^{b} - \theta^{a}\theta^{b} \right)\gamma_{ab}
  +\left(\theta^{a}\rho^{b}-\theta^{b}\rho^{a} \right)F_{ab}
  \right).
\eer{boogie}

Just as for the
tensionless string \cite{lst1}, there are several formulations
that may be derived from \hv{boogie}. Some are easy to analyse from a
geometrical 
point of view, and we now turn to these.  The action (\ref{boogie})
may be written 
\beq
S={\textstyle{1\over 4}}\int d^{p+1}\xi V^iW^j(\g_{ij}+F_{ij}),
\eeq{VWac}
where we identify
\ber
&&V^i\is {1\over {\sqrt \l}}\left( 1, -(\th^a+\r^a)\right),\cr
&&W^i\is {1\over {\sqrt \l}}\left( 1, (\th^a-\r^a)\right).
\eer{VW}
The Lagrange multipliers have been interpreted as vector densities (of
weight $\frac{1}{4}$) to ensure diffeomorphism invariance. Note that the
rank of $V^iW^j$ is $2$, which explains why $p=1$ is a special case
(treated in Section $2$). In what follows, we will be concerned with $p>1$. For
the tensionless fundamental string a formulation analogous to (\ref{VWac}) has
been extensively used, e.g., for constructing spinning models both in components,
\cite{lst2}, and in a superspace formulation \cite{lr}. In the those cases it was
useful to interpret the vector densities as a degenerate vielbeins. We will
elaborate on this interpretation for the present case below.

The degrees of freedom add up when going from (\ref{boogie}) to
(\ref{VWac}), due to the invariance of (\ref{VWac}) under the local
scale transformations
\beq 
V^i{}'= \L (\xi )V^i, \qquad W^i{}'= \L^{-1} (\xi )W^i.
\eeq{scale}
The equations of motion that follow from variation of (\ref{VWac})
with respect to $W^i$, $V^i$, $A_i$ and $X^\mu$ are
\ber
 V^{i}\left(\gamma_{ik} + F_{ik}\right) &=& 0,\cr
 \left(\gamma_{ik} + F_{ik}\right)W^{k} &=& 0,\cr
 \partial_{i}\left(V^{[i}W^{k]}\right) &=& 0,\cr
 \partial_{i}\left(V^{(i}W^{k)}\partial_{k}X^{\mu}\right) &=& 0,
\eer{eom}
(For more clarity and without loss of generality we have put
$G_{\mu\nu}=\eta_{\mu\nu}$).  The first (and second) show that 
$\g_{ij}+F_{ij}$ has
null-eigenvectors. This means that it is degenerate and has zero
determinat. Unlike the usual case \cite{lst1}, this does not mean that the
world volume is a null surface, however. It merely implies a relation
between the $X^\mu$ and the $A_i$ fields. We may think of the dynamics
of the world-volume vector fields as being determined by the
coordinate fields. We will analyse the equations further in the next
section.

Before doing this let us note yet another form for the action bringing
out the geometrical content even better. Defining
\beq
V^i\is E^i_1+E^i_2, \qquad W^i\is E^i_1-E^i_2,
\eeq{Edef}
we rewrite (\ref{VWac}) as described in Section $2$:
$$
S={\textstyle \frac{1}{4}}\int d^{p+1}\xi
(E^i_1E^j_1-E^i_2E^j_2-E^i_{[1}E^j_{2]})(\g_{ij}+F_{ij}).
$$
Integrating out the $A_i$-field we are left with
$$
S={\textstyle{1\over 4}}\int d^{p+1}\xi
\left(E_AX^\mu E_BX^\nu (\eta^{AB}G_{\mu\nu}(X)-\e^{AB}B_{\mu\nu}(X)\right),
$$
where $A,B,..$ are $2D$ Lorentz indices, as described in Section $2$. The
symmetry (\ref{scale}) gets replaced by
$2D$ local Lorentz symmetry in this formulation. In fact, we may view
(\ref{VWac}) as the action (\ref{Eac}) expressed in $2D$ light-cone
coordinates, as will be further discussed below. 

As a final general comment on the geometrical formulation we note that
connections for diffeomorphisms, the scale transformations
(\ref{scale}) or 
local Lorentz
transformations are in general needed for a covariant description of
the 
models. We will not need them here, but note that the analogous
objects were very useful for formulating
a degenerate version of the 2D
supergravity in 2D superspace to describe spinning version of the
tensionless string \cite{lr}.

As an alternative to the above description, (and to make contact with
\cite{blt}), we now integrate out various fields in the action
(\ref{boogie}). The equation of motion for $\theta^{a}$, using the
spacial part of the metric $\hat{\gamma}_{ab} = \gamma_{ab}$, is
\ber
 \theta^{a} = \hat{\gamma}^{ab}\left(F_{0b}
   +\rho^{c}F_{bc}\right),
\eer{thetadied}
which leaves us with
\ber
{\cal L} = \frac{1}{4\lambda}\left( \gamma_{00} - 2\rho^{a}\gamma_{0a}
 + \rho^{a}\rho^{b}\gamma_{ab} +
 \left(F_{0a}+\rho^{c}F_{ac}\right)\hat{\gamma}^{ab}
 \left(F_{0b}+\rho^{d}F_{bd}\right)\right).
\eer{leave}
Continuing to integrate out $\rho^{a}$ we find an equation of motion
\beq
 \rho^{a} = \left(\hat{M}^{-1}\right)^{ab}M_{b0},
\eeq{motion}
where we have introduced the matrix $M$ defined as
$M_{ik} = \gamma_{ik} - F_{ia}\hat{\gamma}^{ab}F_{bk}$ and its spacial part
$\hat{M}_{ab}=M_{ab}$.
Inserting this into the Lagragian we find
\ber
 {\cal L} = \frac{1}{4\lambda}\left( M_{00} 
      - M_{0a}\left(\hat{M}^{-1}\right)^{ab}M_{b0}\right).
\eer{finale}
Now, by introducing the matrices $N_{ik}=\left(\g+F\right)_{ik}$ and
$K_{ik}=\left(\g-F\right)_{ik}$ and their spacial parts $\hat{N}$ and
$\hat{K}$ we can rewrite this as
\ber
 {\cal L} &=&\frac{1}{4\l}\left(
   \g_{00} -F_{0a}\left(\hat{\g}^{-1}\right)^{ab}F_{b0}
 \right. \cr && \left.
  - \left(N_{0a}-F_{0e}\left(\hat{\g}^{-1}\right)^{ef}N_{fa}\right)
  \left(\hat{N}^{-1}\right)^{ab}\g_{bc}
  \right. \cr && \left. \times
  \left(\hat{K}^{-1}\right)^{cd}
  \left(K_{d0}+K_{dg}\left(\hat{\g}^{-1}\right)^{gh}F_{h0}\right)\right),
\eer{mess}
which can be rewritten as
\ber
 {\cal L} &=& \frac{1}{4\l}\left(\g_{00} -
   N_{0a}\left(\hat{N}^{-1}\right)^{ab}F_{b0}
 \right. \cr && \left.
  +F_{0a}\left(\hat{K}^{-1}\right)^{ab}K_{b0}
  -N_{0a}\left(\hat{N}^{-1}\right)^{ab}\g_{bc}\left(\hat{K}^{-1}\right)^{cd}
  K_{d0}\right).
\eer{lessmess}
Using the properties of these matrices under transposition one can
simplify this expression
\beq
 {\cal L} = \frac{1}{4\lambda}\left(
   \g_{00} - N_{0a}\left(\hat{N}^{-1}\right)^{ab}N_{b0}\right),
\eeq{simpler}
wich is the same as
\ber
 {\cal L}= \frac{\det\left(N\right)}{4\lambda\det\left(\hat{N}\right)},
\eer{detM=0}
and by a suitable redefinition of the lagrange multiplier we arrive at
the Lagrangian
\beq
 {\cal L} = V\det\left(\g+F\right),
\eeq{tsey}
which is the tensionless limit of the Born-Infeld action used in
\cite{blt,Tseytlin}.

\section{Solution}
We will now study the solutions to the equations \hv{eom} and show that these
describe a collection of tensile strings.
The first two equations of motion (\ref{eom}) can be reduced to the
equations 
\ber
 V^{i}V^{k}\gamma_{ik} &=& 0,\cr
 W^{i}W^{k}\gamma_{ik} &=& 0,
\eer{sil2}
which say that $V$ and $W$ are null-like vector fields in the induced
metric.
The third equation can be written as $\left[V,W\right]^{i} =
(\partial\cdot W) V^{i} - (\partial\cdot V) W^{i}$, which, after choosing
the gauge $\partial\cdot V = \partial\cdot W = 0$ says that $V$ and $W$
commute and define good coordinates. The coordinates thus defined
coordinatize two dimensional submanifolds of the world volume. They
will be the world sheets of the constituent strings. On each world
sheet we have the differential operators
\ber
 \partial_{+} &=& V^{i}\partial_{i},\cr
 \partial_{-} &=& W^{i}\partial_{i}.
\eer{sil3}
Using these and the gauge choice we see that the equations of motion
reduce to
\ber
 \gamma_{++} &=& 0,\cr
 \gamma_{--} &=& 0,\cr
 \partial_{+}\partial_{-}X^{\mu} &=& 0,
\eer{sil4}
which are exactly the equations of motion for a 
string in light cone gauge. The additional coordinate dependence 
of $X^{\mu}$ now becomes a
label distinguishing between different string world sheets.

For the special case when $V^i$ and $W^i$ are parallel, an analogous
analysis shows that the world volume splits into a collection of
massless particles.

\section{Discussion}

Rederiving the first order action for Born-Infeld theories using a
Hamiltonian approach, we found a direct way of investigating the
strong coupling limit of $D$-branes by formally taking the zero
tension limit. For $p=1$ this gave us a unified description of tensile
and tensionless strings reminicent of that proposed in \cite{blt}, but
here derived as a limit of an underlying description. For $p>1$ an
interesting structure emerged. In this limit the $D$-branes may be
described as composed of tensile strings. The various strings are then
labeled by the additional world-volume coordinates, and the tension
may also depend on those. This limit is described by a novel kind of
action involving a degenerate vielbein wich connects the world-volume
to a $2$D internal space of Lorentzian signature.

This ``parton'' picture might be interesting in connection with the
M-theory matrix model \cite{bfss}, where the partons are $D$ 0-branes,
corresponding to the special case where the vector fields $V^{i}$ and
$W^{i}$ are parallel. Perhaps a more interesting candidate is the
recently proposed matrix model for type IIB string theory \cite{ikkt}.
Since there are no 0-branes in that theory the partons would have
to be strings.

The form of the action seems excellently suited for a superspace
description, and we plan to return to this question of spinning
$D$-branes in the near future.
\bigskip

{\bf Acknowledgements} \vskip .2in
\noindent
We are grateful to Eric Bergshoeff, Bo Sundborg and Paul Townsend for
discussions. The work of UL was supported in part by NFR grant
No. F-AA/FU 04038-312 and by NorFA grant No. 96.55.030-O.

\end{document}